\documentclass[aps,pre,superscriptaddress,showpacs,floatfix,twocolumn]{revtex4}
\usepackage{graphicx}   
\pdfoutput=1
\begin{document}

\title{Circumferential gap propagation in an anisotropic elastic bacterial sacculus}

\author{Swadhin Taneja}	
\email{swadhin.taneja@dal.ca}
\author{Benjamin A. Levitan}	
\author{Andrew D. Rutenberg}	
\email{adr@dal.ca}
\affiliation{Dept. of Physics and Atmospheric Science, Dalhousie University, Halifax, Nova Scotia, Canada B3H 4R2}

\date{\today}
\begin{abstract}  
We have modelled stress concentration around small gaps in anisotropic elastic sheets, corresponding to the peptidoglycan sacculus of bacterial cells, under loading corresponding to the effects of turgor pressure in rod-shaped bacteria. We find that under normal conditions the stress concentration is insufficient to mechanically rupture bacteria, even for gaps up to a micron in length. We then explored the  effects of stress-dependent smart-autolysins, as hypothesised by Arthur L Koch  [Advances in Microbial Physiology 24, 301 (1983); Research in Microbiology 141, 529 (1990)]. We show that the measured anisotropic elasticity of the PG sacculus can lead to stable circumferential propagation of small gaps in the sacculus. This is consistent with the recent observation of circumferential propagation of PG-associated MreB patches in rod-shaped bacteria. We also find a  bistable regime of both circumferential and axial gap propagation, which agrees with behavior reported in cytoskeletal mutants of {\em B. subtilis}.  We conclude that the elastic anisotropies of a bacterial sacculus, as characterised experimentally, may be relevant for maintaining rod-shaped bacterial growth. 
\end{abstract}
\maketitle

\section{Introduction}
Outside the inner membrane (IM) of bacteria, peptidoglycan forms a covalently-linked mesh that accommodates continual growth while preserving cell-shape and integrity.   The peptidoglycan (PG) sacculus of rod-shaped Gram-negative (G-) bacteria such as {\em Escherichia coli} is best understood, and is thought to be composed of approximately one layer of circumferential glycan strands connected by peptide cross-links \cite{Holtje1998, Vollmer2010, Typas2012}. This thin mesh retains significant osmotic (turgor) pressures, ranging from $10-30$ kPa in growth medium to $400$ kPa in water \cite{Yao2002, Deng2011}.

There is not yet a compelling model of how rod-shaped bacteria grow longer without growing wider --- in order to accommodate a doubling of length before midcell division.  The challenge is that the microscopic structure of the sacculus is disordered, with a broad range of glycan chain lengths that are much shorter than the bacterial circumference \cite{Holtje1998, Vollmer2010} and significant disorder in their alignment visible under cryo-electronmicroscopy (cryo-EM) \cite{Gan2008}.  This appears to rule out straightforward templating mechanisms in which glycan hoops are copied to accommodate extension (see e.g. \cite{Holtje1998}).  Nevertheless, recent experiments in both Gram-positive (G+) \cite{Garner2011, DominguezEscobar2011} and G- \cite{Teefelen2011} bacteria indicates that PG synthesis drives local MreB patches circumferentially around rod-shaped bacteria.  While such circumferential synthesis can explain rod-like elongation, the mechanism of the circumferential orientation remains unknown.  

We would expect that local disorder, due to distributions of glycan chain lengths \cite{Holtje1998, Vollmer2010} and to variations in glycan chain orientation \cite{Gan2008}, leads to mesoscopic gaps in the PG sacculus. Indeed, gaps of various diameters up to approximately $20$ nm have recently been reported in {\em E. coli} \cite{Turner2013}. From continuum elastic theory \cite{Sadd2005}, we expect stress concentration around these gaps in the PG sacculus.  

Rod-shaped bacteria have strikingly anisotropic elasticity and loading. Atomic-force microscopy (AFM) of G- bacteria consistently obtains an approximately two-fold anisotropy in elastic constants \cite{Yao1999, Deng2011}, with the circumferential (glycan-chain) direction stiffer than the axial (peptide-bond) direction. There is also a two-fold anisotropy in the tension within the membrane of rod shaped cells due to their geometry combined with turgor pressure --- with twice the tension in the circumferential direction \cite{Koch1983}.  We expect that the anisotropic elasticity and loading would affect the stress concentration around gaps in the PG sacculus. 

In this paper we characterise necessary conditions for stable gap propagation due to stress-concentration within the context of an anisotropic elastic model of the PG sacculus. We find that stress concentration around observed gap sizes is not sufficient to mechanically rupture the PG with normal turgor pressures, and that large gap sizes will mechanically rupture the membrane \cite{Daly2011} before the PG. Nevertheless, the stress-dependent ``smart autolysins'' proposed by Koch \cite{Koch1983, Koch1990} could exploit the stress-concentration around PG gaps to propagate gaps without mechanical rupture.  Similarly, strain-dependent modulation of PG synthesis through the action of outer membrane lipoproteins has also been proposed \cite{Typas2012}.  Our results show that stress concentration around small gaps in the anisotropic PG sacculus could both localize and orient stress (or strain)-sensitive PG degradation and synthesis.  We believe that this localization and orientation may explain the circumferential MreB patch motion reported experimentally \cite{Garner2011, DominguezEscobar2011, Teefelen2011}. 

\section{Model}
\begin{figure}[t*]
	\includegraphics[width=0.5\linewidth, angle=0]{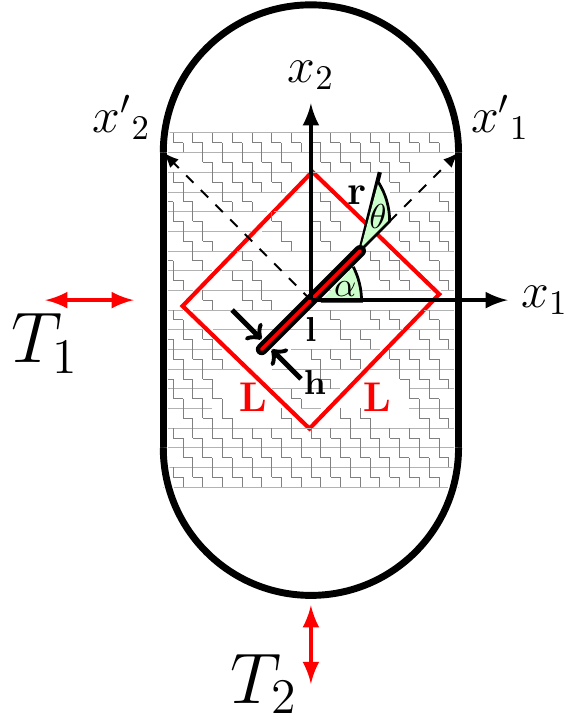}
\caption{(color online) The PG sacculus of a rod-shaped bacterial cell is illustrated, with membrane tensions $T_1$ and $T_2$ in the indicated directions. A traction-free gap of length $l$ and width $h$,  at an inclination  $\alpha$ with respect to the circumferential directions is illustrated (not to scale). We solve the elastic stress field within a small square patch of size $L \times L$, as indicated, oriented with the gap. A possible gap extension angle $\theta$, with respect to the gap inclination, is also illustrated.}
\label{f:crackgeo}
\end{figure}

For small regions of the bacterial surface, as illustrated in Fig.~\ref{f:crackgeo}, we assume a linear, Hookean, relation between stress and strain:  
\begin{equation}
	 \sigma_{ij} = C_{ijkl} \epsilon_{kl},\;\;
	    \epsilon_{ij} = \frac{1}{2} (u_{i,j}+u_{j,i}),\;\;
	i,j,k,l=1,2
\label{e:srst}
\end{equation} 
where $\sigma$, $\epsilon$  and $C$ are the stress, strain and stiffness tensors, respectively and $u$ is the local (two-dimensional) displacement vector. Partial derivatives are indicated by commas in $u_{i,j}$. Symmetries of the stress and strain tensors imply $C_{ijkl}=C_{jikl}=C_{ijlk}=C_{klij}$, where we also exploit the work-conjugate nature of stress and strain \cite{Sadd2005}.  Restricting ourselves to an orthotropic material for simplicity, with reflection symmetry around $x_1$ and $x_2$, further reduces the independent stiffness components to only four: $C_{1111} = E_1/(1-\nu_{12}\nu_{21})$,   $C_{2222} = E_2/(1-\nu_{12}\nu_{21})$,  $C_{1122} =C_{2211} = \nu_{12}E_1/(1-\nu_{12}\nu_{21})=  \nu_{21} E_2/(1-\nu_{12} \nu_{21})$, and $C_{1212} \equiv G$ --- where $E_i$ is the Young's modulus in direction $i$, $\nu_{ij}$ is the Poisson's ratio that corresponds to contraction in direction $j$ when extension is applied in $i$; and $G$ is the shear modulus. 

We define the geometric mean $\nu \equiv \sqrt{\nu_{12}\nu_{21}}$, and then fix the shear modulus $G$ using the ratio $C_{1212}/C_{1122} = (1-\nu)/(2\nu)$ seen for isotropic systems \cite{Sadd2005}. Similarly, we define the geometric-mean of the anisotropic elastic constants, $E \equiv \sqrt{E_1 E_2}$ and the ratio $\eta \equiv E_1/E_2$, and scale all stiffness components and stresses globally by $C_{2222}$. We have $C_{1111}/C_{2222} = \eta$ and  $C_{1122}/C_{2222}=C_{2211}/C_{2222} = \nu \sqrt{\eta}$.  As the Poisson ratio has not been measured for bacterial surfaces, we use a rubber-like $\nu=1/2$ --- and check that varying $\nu$ within a two-fold range does not significantly change our results. We systematically vary the ratio of elastic constants $\eta$, as well as the scaled gap length $l/h$. 

As illustrated in Fig.~\ref{f:crackgeo}, a traction vector $\vec{T}$ will act on the boundaries of small regions of the bacterial surface with $T_{1}=2 T_{2}$, due to the cylindrical geometry and turgor pressure, where $x_1$ and $x_2$ are the circumferential and axial directions, respectively.  We introduce a traction-free gap, of length $l$ and width $h$ and inclined at angle $\alpha$ with respect to $x_1$, as indicated. The gap terminates in semicircular caps, of radius $h/2$, so that $l \geq h$. Since $h>0$, stress will be concentrated at the gap tip but is not singular there. The gap is aligned within an $L \times L$ patch for computational convenience.

To determine the stress and strain fields everywhere in the patch we computationally solve the mechanical-equilibrium equation
\begin{equation}
	\sigma_{ij,j} = 0,
	\label{e:eqm}
\end{equation}
using PDE2D (ver. 9.3) \cite{Sewell1993} with an adaptive triangular grid. To minimize finite-size effects, we use a linear patch size $L = 10,000 h$  and to minimize discretization effects we use  $N = 100,000$ triangles.  We show results for $l/h=2$, unless otherwise noted. 

From the stress-field solution of Eqn.~\ref{e:eqm}, we consider the circumferential stress $\sigma'_{\theta \theta}$ at the tip of the gap, where the prime indicates that $\theta$ (and the stress) is measured with respect to the gap orientation $\alpha$.  This stress can mechanically break bonds, or could couple to local PG degradation through e.g. hypothesized smart autolysins \cite{Koch1983, Koch1990}, and so lead to gap extension.  We assume an orthotropic toughness, or yield stress, of the sacculus, $K'_C(\theta)=K_1 \cos^2(\theta+\alpha) + K_2 \sin^2(\theta+\alpha)$, where $K_1$ is the toughness with respect to circumferential gap extension and $K_2$ with respect to axial gap extension.  The yield stress would also characterize the stress at which smart autolysins begin to rapidly break bonds. The anisotropy of $K'_C$ is $\kappa \equiv K_2/K_1$. 

For a given $\alpha$, the gap will propagate at the angle $\theta_{max}$ that maximizes the ratio between stress and toughness (see, e.g., \cite{Carloni2005, Buczek1985}),
\begin{equation}
	R(\theta) \equiv \frac{\sigma'_{\theta\theta}}{K'_C(\theta)/\sqrt{2\pi r}},
	\label{e:ccritn}
\end{equation}
where the $1/\sqrt{r}$ factor captures the expected dependence of the stress enhancement near the gap tip. We evaluate $R$  at the gap tip, with $r=h/2$. Since we are considering linear elasticity, the angle at which $R$ is maximized is independent of the turgor pressure, or the magnitudes of the elasticity or toughness. Nevertheless, the various anisotropy factors ($\eta$, $\kappa$, and $T_1/T_2=2$) remain important. 

We identify gap inclinations $\alpha_0$ that will continue to propagate straight, i.e. where $\theta_{max}=0$. We would also like gap propagation to be stable to perturbations arising from, e.g., local disorder. Our stability condition is that gaps  with $\alpha \lesssim \alpha_0$ will have $\theta_{max}>0$ and gaps with $\alpha \gtrsim \alpha_0$ will have $\theta_{max}<0$.   Inclinations with $\theta_{max}=0^\circ$ that do not satisfy our stability condition are unstable. 

\section{Results}

\begin{figure}[t*]
\includegraphics[width=\linewidth]{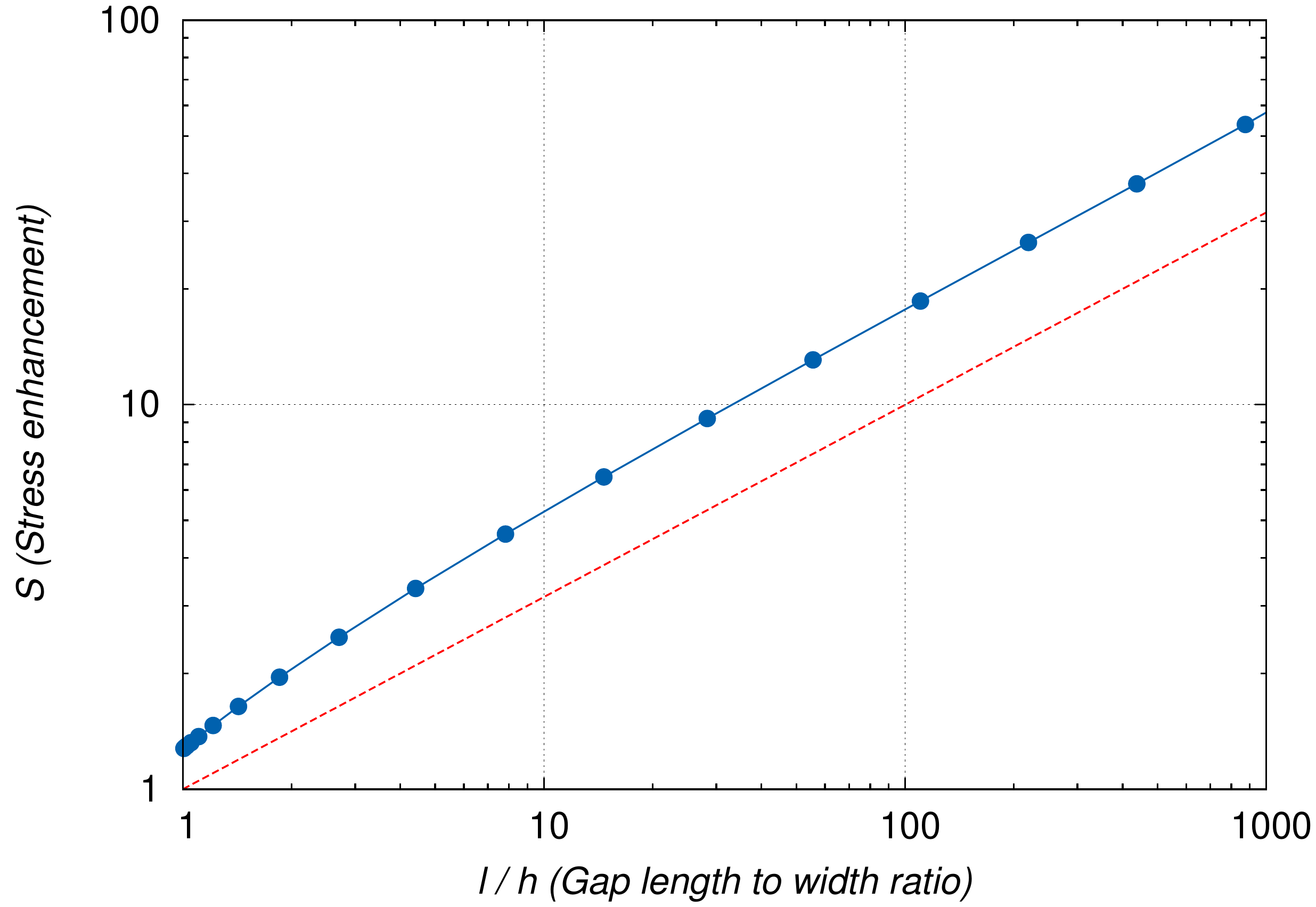}
\caption{(color online) Stress enhancement ratio, $S=\sigma'_{\theta \theta}/T_2$, perpendicular to the tip of a circumferential gap (with $\theta=\alpha=0^\circ$), for various gap lengths is indicated by the solid blue circles and with a solid line as a guide to the eye. $S>1$ reflects stress concentration due to the traction-free gap. A dashed red line shows the asymptotic $\sqrt{l/h}$ behavior expected in the thin crack limit at large $l/h$. The  stiffness ratio $\eta =2$.} 
\label{f:length}
\end{figure}

We first consider the magnitude of the stress enhancement at the gap tip. In Fig.~\ref{f:length} we show the stress enhancement as a function of the ratio $l/h$ for a stable circumferential gap inclined at $\alpha=0^\circ$ with $\eta=2$. The stress enhancement is the ratio of the tip stress to the axial stress at the system boundary, i.e. ${\sigma'_{\theta \theta}}/T_2$ with $\theta=0^\circ$. At larger $l/h$ we obtain the expected asymptotic $\sqrt{l/h}$ scaling of stress enhancement with gap length \cite{Sadd2005}.  

While the stress enhancement can be considerable for long gaps, we do not expect frequent mechanical rupture of covalent bonds occur during normal PG growth.  Given a characteristic covalent bond strength of $1$ nN \cite{Grandbois1999} and a PG bond spacing of $4$ nm for G- bacteria \cite{Demchick1996}, then a local stress of $0.25$ nN/nm is needed to quickly rupture bonds. For a  turgor pressure of $30$ kPa \cite{Deng2011}, bacterial radius $R=0.5 \mu$m, and the stress concentrations in Fig.~\ref{f:length}, this corresponds to $l/h \gtrsim 300$. If $h \approx 4$ nm, then this requires gap lengths longer than $1 \mu$m.  This is much longer than the patches seen in MreB localization experiments \cite{DominguezEscobar2011, Garner2011, Teefelen2011}, and also much larger than the critical circular gap radius of $20$ nm for membrane instabilities \cite{Daly2011}. Mechanical  bond rupture is an unlikely mechanism for coupling stress with anisotropic gap propagation during normal growth. 

Alternatively,  Koch proposed \cite{Koch1983, Koch1990} ``smart'' or stress-dependent autolysins that could couple stress to enzymatic bond degradation. The observation of circumferential patch propagation \cite{DominguezEscobar2011, Garner2011, Teefelen2011} then raises the question of whether stress-dependent enzymatic bond degradation could lead to the observed circumferential propagation. Accordingly, we have investigated the orientation of stable gap propagation within our anisotropic elastic model.  Interestingly, the results do not qualitatively vary on the gap length $l/h$ once $l/h>1$. We show results for $l/h=2$, corresponding to relatively small gaps. 
 
\begin{figure}[t*]
	\includegraphics[width=\linewidth]{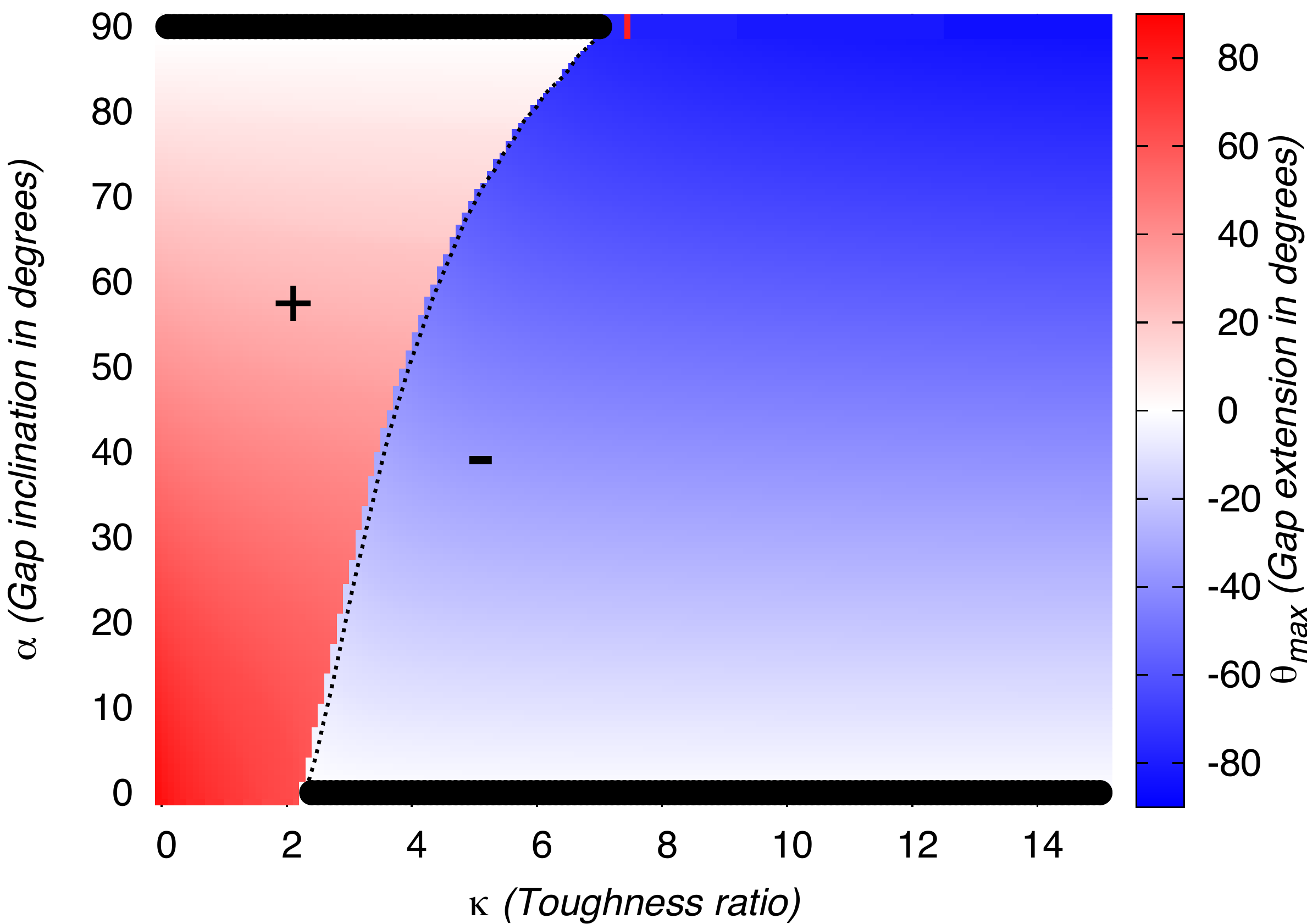}
\caption{(color online) Bifurcation diagram of stable gap directions. Solid and dotted black lines indicate stable and unstable gap inclination directions $\alpha_0$, respectively, vs. the PG toughness ratio $\kappa \equiv K_2/K_1$.  Colors indicate inclinations $\alpha$ with expected deviations in the $\theta_{max} >0$ (red, ``$+$'') or $\theta_{max}<0$ (blue, ``$-$'') directions --- according to the scale at the right. The  stiffness ratio $\eta =2$. Circumferential gaps have $\alpha=0^\circ$, while longitudinal gaps have $\alpha=90^\circ$, as indicated in Fig.~1.}
\label{f:alphakykx} 
\end{figure}

In Fig.~\ref{f:alphakykx} we plot $\theta_{max}$  values for a variety of gap inclinations $\alpha$ and toughness anisotropies $\kappa$, all for a stiffness ratio $\eta=2$ \cite{Yao1999, Deng2011}.  Regions of positive and negative $\theta_{max}$ are in red (with ``+'') and blue (with ``-''), respectively, with $\theta_{max}=0^\circ$ shown in white.  All angles are reported in degrees for convenience. The lines of $\theta_{max}=0^\circ$ correspond to either unstable (dotted line) or stable (solid line) gap propagation directions. We expect that PG gaps will propagate in stable directions.  Smaller toughness anisotropies, with $\kappa \lesssim 2$, lead to stable axial propagation with $\alpha_0=90 ^\circ$.  Larger toughness anisotropies, with $\kappa \gtrsim 7$, lead to circumferential propagation with $\alpha_0=0^\circ$. While the limiting behavior for $\kappa \rightarrow 0$ and $\kappa \rightarrow \infty$ could be expected, the absence of stable intermediate values of $\alpha$ is surprising. Instead, for $2 \lesssim \kappa \lesssim 7$ we find a line of unstable angles at intermediate $\alpha$ that determine a bistable (coexistence) region in which both axial and circumferential gap propagation is expected. 
                    
\begin{figure}[t*]
\includegraphics[width=\linewidth]{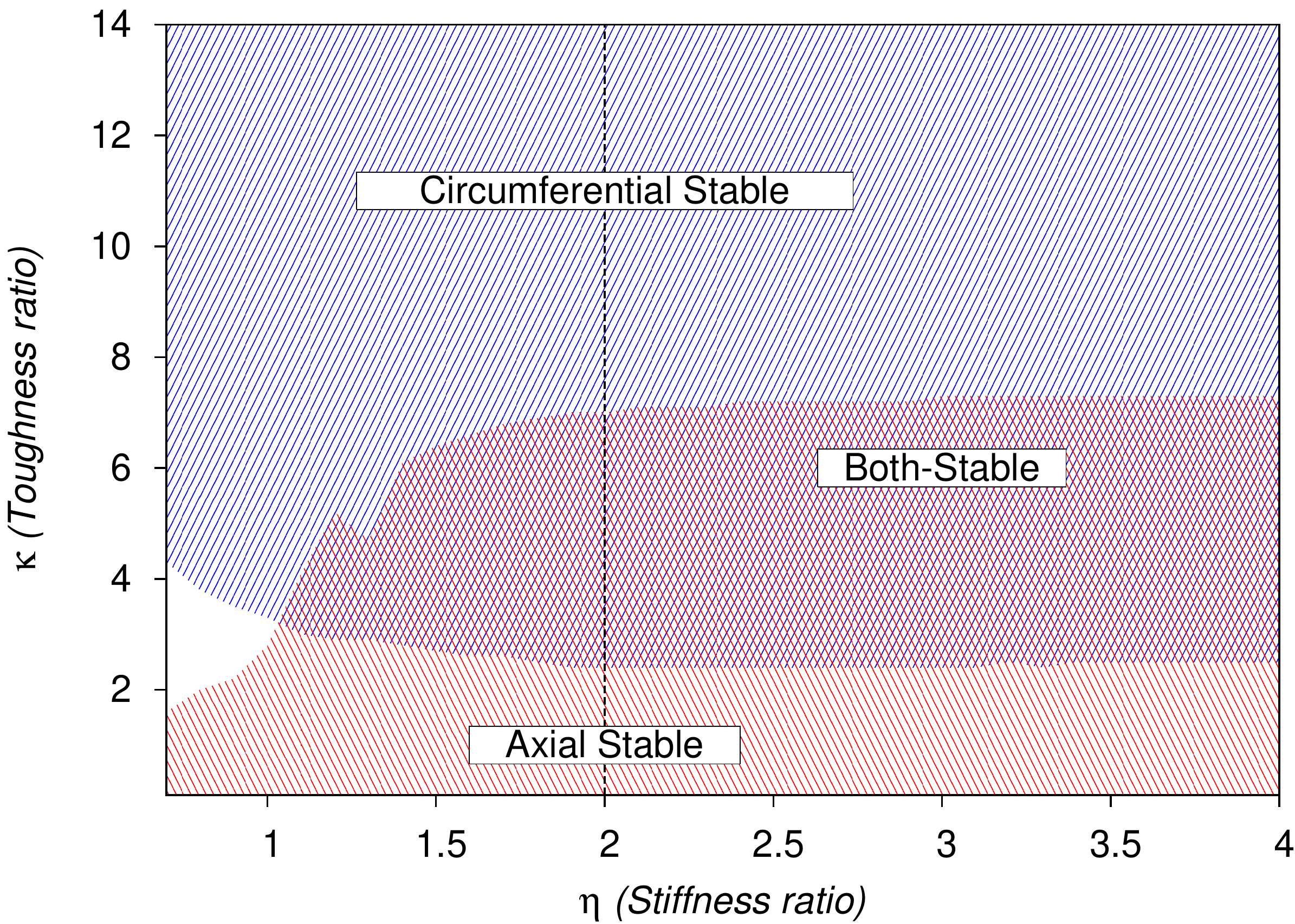}
\caption{(color online) Stability in the space of toughness ratio, $\kappa$, and stiffness ratio, $\eta$. Regions of stable gap inclinations are as labeled, with circumferentially stable gap inclinations ($\alpha_0=0^\circ$) shown in blue, axially stable gap inclinations ($\alpha_0=90^\circ$) shown in red, and regions with bistability in both colors as indicated. A small regime with intermediate stable gap inclinations  ($0^\circ<\alpha_0<90^\circ$) is represented by the uncolored (un-shaded) region for $\eta \lesssim 1$. The vertical dashed line indicates $\eta=2$, appropriate for G- bacteria and corresponding to Fig.~3.} 
\label{f:ExEykykx}   
\end{figure}

In Fig.~\ref{f:ExEykykx} we explore how our bifurcation diagram depends upon the elastic anisotropy $\eta$. Colors indicate stable values of $\alpha_0$, as we vary both $\eta$ and $\kappa$. The vertical dashed line indicates $\eta=2$, as used in Fig.~\ref{f:alphakykx} and as measured in AFM experiments on rod-shaped G- bacteria \cite{Yao1999, Deng2011}. We note that $\eta$ has not yet been measured for G+ bacteria. For all $\eta \gtrsim 1$, we see qualitatively similar behavior --- with stable axial propagation for $\kappa \lesssim 2$, stable circumferential propagation for sufficiently large $\kappa > \kappa_c$, and coexistence for intermediate values of $\kappa$.  The region of circumferential stability is observed for all $\kappa \gtrsim 2$ and is approximately independent of $\eta \gtrsim 1$. The region of axial stability grows significantly for $\eta \lesssim 2$.   The result is that the bistable region grows for $ 1 \lesssim \eta \lesssim 2$, and is approximately independent of $\eta \gtrsim 2$.  For $\eta<1$ there is no bistable region and the circumferential and axial stable regions are separated by a region of stable gap propagation at intermediate angles $0^\circ<\alpha_0 < 90^\circ$. We note that while $\eta=1$ corresponds to isotropic elasticity, the stress is still anisotropic, with $T_1/T_2=2$, due to the overall rod-shaped bacterial geometry.  For isotropic toughness and isotropic elasticity ($\kappa=\eta=1$) Fig.~\ref{f:ExEykykx} indicates that the anisotropic stress leads to stable axial gap propagation, as is commonly observed in cooked sausages.

\section{Summary and discussion}

We have modelled stress concentration around small gaps in anisotropic elastic sheets (see Fig.~\ref{f:crackgeo}), corresponding to the PG sacculus of bacterial cells, under loading corresponding to the effects of turgor pressure in rod-shaped bacteria.  We have explored the hypothesis of stress-dependent extension of gaps, either through mechanical rupture of bonds of the PG sacculus or through additional biological components such as the hypothesized smart-autolysins of Koch \cite{Koch1983, Koch1990}.

We have found that while mechanical rupture of bonds under normal conditions is unlikely without long narrow gaps at least a micron in length (see Fig.~\ref{f:length}),  anisotropies in loading ($T_1/T_2$), elasticity ($\eta \equiv E_1/E_2$), and toughness ($\kappa \equiv K_2/K_1$) can lead a broad regime of stable circumferential gap propagation adjacent to a regime of bistable propagation in which both circumferential and longitudinal gap propagation co-exist (see Figs.~\ref{f:alphakykx} and \ref{f:ExEykykx}). 

It is reassuring that mechanical bond-rupture is not expected under normal conditions, since it would be expected to continue even when cell growth halts.  Nevertheless, during sonication and other externally applied stress-induced rupture of the PG sacculus, stable circumferential tears are seen in both G- and G+ bacteria \cite{sonicate}.  While it is tempting to ascribe these to the stable circumferential gap propagation described in this paper, membrane rupture during sonication is thought to be predominately due to shear \cite{shear} rather than to the hydrostatic pressure treated here.  It would be interesting to see what elastic parameters lead to stable circumferential gap propagation under shear stresses, but we leave that for future work.

We have largely focused on stably oriented gap propagation in the PG sacculus. For $\eta=2$ appropriate for G- bacteria, Fig.~\ref{f:alphakykx} shows that stable circumferential propagation is expected for large but finite values of $\kappa$. We believe that the circumferential propagation of patches of MreB isoforms in rod-shaped G+ \cite{DominguezEscobar2011, Garner2011} and G- bacteria \cite{Teefelen2011} may be explained by circumferential gap propagation in an anisotropically stressed anisotropic elastic medium. Furthermore, the observation of coexistence of axial and circumferential Mbl-GFP patches seen in $\Delta mreb \Delta mbl$ mutants \cite{DominguezEscobar2011} may be explained by the similar coexistence found in our model for slightly smaller $\kappa$, as shown in Fig.~\ref{f:ExEykykx}. Elastic anisotropy has not been directly measured in G+ bacteria so, from Fig.~\ref{f:ExEykykx}, we expect that $\eta \gtrsim 1$, in order to explain the observed bi-directional propagation.   Elasticity studies have shown that MreB can contribute up to $30\%$ of the stiffness of the cell wall \cite{Wang2010}.  This may indicate that wild-type growth operates just above the bistable region in Fig.~\ref{f:ExEykykx}.

While stress-sensitivity has been observed in, e.g., collagen degradation \cite{Adhikari2011} or cleavage of von Willebrand factor \cite{Zhang2009}, it has not been reported for PG degradation.  Nevertheless,  PG hydrolases required for cell growth have only very recently been identified \cite{Singh2012} --- and may now allow more focused examination of stress-sensitivity in their activity.  It also remains to be addressed how PG synthesis could be directed behind gaps to prevent propagating gaps from lengthening. The recent discovery of outer membrane (OM) lipoproteins that activate cell-wall synthesis  \cite{OMlipoprotein} provides a possible mechanism to couple stress and synthesis, at least for G- bacteria.  Gap size will determine a vertical bulge of the bacterial IM \cite{Daly2011}, which could then activate cell-wall synthesis through a transient kissing action of OM lipoproteins.  This ``hernia'' mechanism, first proposed in 1993 \cite{Norris1993} may also explain the patchy lipoprotein localization patterns seen in experiment \cite{OMlipoprotein}.   

The coupling of mechanical forces and biological activity, generally called mechanobiology, can occur in many ways. Recent experimental studies of externally stressed growing bacteria \cite{Amir2013} indicate that at least some mechanical modulation of cell-wall synthesis is taking place.  Our model highlights the potential importance of anisotropic elastic properties of the bacterial sacculus in the growth of rod-shaped bacteria. While the bacterial sacculus is an anisotropic material under anisotropic load, and exhibits anisotropic patterns of growth \cite{DominguezEscobar2011, Garner2011, Teefelen2011} and fracture \cite{sonicate}, only one anisotropic parameter ($\eta \approx 2$ \cite{Yao1999, Deng2011}) has been measured, and only for G- bacteria.  We anticipate that the shear modulus $G$, Poisson's ratio $\nu$, as well as non-orthotropic elastic parameters may affect our results, and therefore could be relevant to bacterial biology.  A key parameter governing the stable directions of gap propagation is the toughness ratio $\kappa$, where larger values leads to circumferential propagation.   However, we should interpret $\kappa$ as a phenomenological parameter reflecting the effective coupling between anisotropic stress and local PG disruption, which may be challenging to measure directly.

There have been two recent modelling approaches to rod-shaped bacterial elongation that can be usefully contrasted with our model. The most recent is the dislocation-driven model of Amir and Nelson \cite{Amir2012}.  Like us, they invoke the observed peptidoglycan disorder to infer the existence of defects. Their dislocations are microscopic dislocations, corresponding to the ends of glycan chains, and they interact with each other through elastic stress fields. They are primarily interested in the interactions between multiple defects as they move  around the sacculus. For simplicity, they work with an isotropic elastic model of the sacculus and build circumferential mobility of defects explicitly into their model. In contrast, our gaps are mesoscopic and as $h \rightarrow 0$ corresponds to two bound dislocations. We ignore interactions between gaps, but explicitly include measured anisotropic elasticity  and so infer circumferential mobility.  Neither Amir and Nelson, nor us, include disorder in the elastic properties of the sacculus. In contrast, the microscopic spring approach of Huang {\em et al} \cite{Huang2008} has a self-consistent model of static but disordered peptidoglycan structure. This was extended \cite{Furchtgott2011} to treat sacculus growth but with explicit circumferential glycan insertion. 

Our approach complements both the dislocation model of Amir and Nelson and the spring model of Huang {\em et al}. We show how an explicit anisotropic elasticity might lead to circumferential propagation, while this circumferential propagation is assumed in the other approaches \cite{Amir2012, Furchtgott2011}.  However, we focus on stress concentration around small gaps in the sacculus while Amir and Nelson consider dislocations.   While we expect small gaps to exist in the sacculus to allow for periplasmic structures such as flagella \cite{Scheurwater2011}, and they appear unavoidable with observed glycan strand-length distributions (see e.g. \cite{Huang2008}), the important experimental question is whether gaps correlate with circumferentially propagating MreB patches \cite{DominguezEscobar2011, Garner2011, Teefelen2011}. If so, this would support our gap extension picture. The emerging field of correlative microscopy \cite{Zhang2013}, combining fluorescence microscopy with subsequent electron-tomography, may be able to resolve this question. 

Since we have no microscopic disorder in our model, beyond the explicit introduction of gaps in the sacculus, we do not recover any dispersion of propagation directions around stable directions. As such, we cannot compare with  experimentally determined dispersions \cite{DominguezEscobar2011, Garner2011, Teefelen2011}. Nevertheless, we feel that stable propagation is required for consistent orientation in the face of disorder. We capture both a stable circumferential regime and an adjacent bistable circumferential/longitudinal regime that we feel may explain the observed stable MreB patch propagation in both G+ \cite{DominguezEscobar2011, Garner2011} and G- \cite{Teefelen2011} bacteria.    We propose that the MreB patches  are associated with small gaps in the PG, and are steered by stress-concentration in this anisotropic elastic system.  


\acknowledgments
We thank Manfred Jericho and Laurent Kreplak for discussions, the Natural Sciences and Engineering Research Council (NSERC) for operating grant support, and the Atlantic Computational Excellence Network (ACEnet) for computational resources. We thank G. Sewell for help with PDE2D. 


\end{document}